\documentclass[aps,showpacs,floatfix,11pt,preprintnumbers]{revtex4}
\usepackage{psfig}
\begin{document}
\preprint{LBNL-58722}
\title{$\alpha \simeq \pi/2$ from \\ supersymmetric spontaneous flavor breaking}
\author{Javier Ferrandis}
\email{ferrandis@mac.com}
\homepage{http://homepage.mac.com/ferrandis}
\affiliation{MEC postdoctoral fellow at  the Theoretical Physics Group \\
 Lawrence Berkeley National Laboratory  \\
One Cyclotron Road, Berkeley CA 94720}
\begin{abstract}
We propose a flavor model where both CP and flavor symmetries are broken at the supersymmetric level.
The model is an effective SU(5) theory based on a U(2) horizontal symmetry. 
The minimum of the supersymmetric scalar potential can be exactly solved to yield 
a realistic pattern of charged fermion masses.
The Higgs sector contains a symmetric, an antisymmetric and two vector fields, plus their U(2) conjugates.
These Higgs fields are the only fields strictly required to break the flavor and CP symmetries
and generate masses for all charged fermions including the up-quark.
The model predicts the existence of an absolute minimum in the space of CP-phases. 
The value  $\alpha \simeq \pi/2$ is predicted in a particular limit of the parameter space
of the model. 
\end{abstract}
\maketitle
\newpage
%
\section{Introduction}
The flavor problem is a persistent challenge of modern particle physics. 
Despite the numerous and diverse ideas have been proposed to explain the fermion mass
hierarchies there are only a few models that can be considered realistic, 
{\it i.e.} that fit the data with precision 
\cite{Barbieri:1995uv,Barbieri:1996ww,Barbieri:1997tu,babu,ross,dermiraby,chenmahan,albarr,Kim:2004vk,aracar}. 
Some of these models
are afflicted by serious limitations: they are effective field theories,
have to resort to contrived flavor symmetries,
or introduce the breaking of the flavor symmetry {\it ad-hoc} 
({\it i.e.}, hierarchies are generated through Higgs vevs that are postulated 
rather than obtained from the minimization of the scalar potential). 
Underlying these difficulties 
is the fact that the unknown flavor symmetry, active at some
high energy scale, is completely broken at low energies,
with the result that
more than one Higgs doublet becomes necessary. 
Furthermore, to generate a realistic spectra of fermion masses,
Higgs fields playing no role in the breaking of the flavor symmetry usually have to be introduced.
In many cases, once the Higgs sector has been postulated, the analysis of the minimum 
becomes intractable, since the number of couplings allowed by 
the flavor symmetry rapidly increases with the number of Higgses. 

One would like to simplify the analysis of the flavor vacuum by reducing, as much as possible,
the number of couplings relevant in the breaking of the flavor symmetry. 
In this paper, we propose a realistic three-generation supersymmetric flavor model where 
the spontaneous breaking of 
CP and flavor symmetries occur at the supersymmetric level,
obviating supersymmetry breaking terms.
We have introduced the minimum number of Higgs fields required to break
flavor and CP symmetries and generate a realistic pattern of charged fermion masses.
In this model, supersymmetry allow us not only to reduce the number
of parameters in the flavor sector, but also to solve 
the flavor vacuum analytically. The result is a more predictive flavor model. 
\section{The model \label{model}}
Let us consider a supersymmetric SU(5) model based on a horizontal U(2) flavor symmetry.
We will assume that left and right handed 
third generation matter fields unify in 
the usual representations $10$ and $\bar{5}$ of SU(5)
\begin{equation}
\psi_{10}^3 , 
\quad
\psi_{\bar{5}}^3,
\end{equation}
which transform as singlets of U(2). We will assume that first and second generations
transform as fundamental representations of U(2), which we will denote as
\begin{equation}
\Psi_{10} =
\left(
\begin{array}{c}
\psi_{10}^1 \\
\psi_{10}^2
\end{array}
\right),
\quad
\Psi_{\bar{5}} =
\left(
\begin{array}{c}
\psi_{\bar{5}}^1 \\
\psi_{\bar{5}}^2
\end{array}
\right).
\end{equation}
We will assume that there are two U(2) singlet Higgs fields transforming under
the representations $5$ and $\bar{5}$ of SU(5), ${\cal H}_{5}$ and 
${\cal H}_{\bar{5}}$, which contain the usual electroweak symmetry breaking Higgs fields.
Finally we will assume that there are four flavor breaking chiral superfields 
\begin{equation}
{\cal S}^{ab}, \quad 
{\cal A}^{ab}, \quad
{\cal F}_1^a, \quad
{\cal F}_2^a,
\end{equation}
and their U(2) conjugates, $\bar{{\cal S}}$, $\bar{{\cal A}}$, $\bar{{\cal F}}_1$, $\bar{{\cal F}}_2$.
These fields transform as a symmetric, antisymmetric and vector fields under U(2). 
Therefore at the renormalizable level only two Yukawa interactions are allowed by the symmetry
\begin{equation}
\frac{1}{4} h_t \psi_{10}\psi_{10} {\cal H}_5 +
\sqrt{2} h_b \psi_{10}\psi_{\bar{5}} {\cal H}_{\bar{5}}.
\end{equation}
These terms will generate masses for the third generation fermions predicting
$m_b=m_{\tau}$ at the GUT scale. The supersymmetry conserving scalar potential
is determined by the superpotential of the model. The most general renormalizable
superpotential invariant under U(2) can be separated in three parts, 
${\cal W} = {\cal W}_A + {\cal W}_S + {\cal W}_F$. ${\cal W}_S$ contains
all the interactions which include the symmetric tensors ${\cal S}$ and $\bar{{\cal S}}$
\begin{eqnarray}
{\cal W}_S &=& \mu_S {\cal S} \bar{\cal S} +
\lambda_1 ( {\cal F}_1 \bar{\cal S} {\cal F}_1 + \bar{\cal F}_1 {\cal S} \bar{\cal F}_1 ) + 
\nonumber \\
&& \lambda_2 ( {\cal F}_2 \bar{\cal S} {\cal F}_2 + \bar{\cal F}_2 {\cal S} \bar{\cal F}_2 ) + 
 \lambda_s ( {\cal F}_1 \bar{\cal S} {\cal F}_2 + \bar{\cal F}_1 {\cal S} \bar{\cal F}_2 ) .~~~
\end{eqnarray}
${\cal W}_A$ contains
all the interactions which include the antisymmetric tensors ${\cal A}$ and $\bar{{\cal A}}$,
\begin{equation}
{\cal W}_A = \mu_A {\cal A} \bar{\cal A} +
\lambda_a ( {\cal F}_1 \bar{\cal A} {\cal F}_2 + \bar{\cal F}_1 {\cal A} \bar{\cal F}_2 ),
\end{equation}
and ${\cal W}_F$ contains
all the interactions which include only the vector fields ${\cal F}_1,{\cal F}_2$ and 
$\bar{{\cal F}}_1,\bar{\cal F}_2$,
\begin{equation}
{\cal W}_F = \rho_1 {\cal F}_1 \bar{\cal F}_1 +
\rho_2 {\cal F}_2 \bar{\cal F}_2 +
\rho_{3}  {\cal F}_1 \bar{\cal F}_2 +  \rho_{4}  {\cal F}_2 \bar{\cal F}_1 .
\end{equation}
We note that the fields ${\cal F}_{1}$ and ${\cal F}_{2}$ have the same quantum numbers.
We could rewrite the superpotential as a function of two alternative  
vector fields, mixings of the original vector fields.
Therefore we will assume from now on, without any loss of generality,
that the fields fields ${\cal F}_{1}$ and ${\cal F}_{2}$  have been conveniently 
defined to cancel the couplings $\lambda_{1}$ and $\lambda_{2}$. 
This choice will considerably simplify the forecoming analysis.
In a supersymmetric theory the vacuum energy is bound to vanish. Therefore the supersymmetric
scalar potential associated to ${\cal W}$, ${\cal V} = {\cal V}_S + {\cal V}_A + {\cal V}_F$,
must be zero at the minimum. If the real couplings $\mu_S$ and $\mu_A$
are not zero there is a solution satisfying ${\cal V}_S = 0 ={\cal V}_A$. 
This can be given by
\begin{eqnarray}
\left< {\cal S}^{ab} \right> &=& \frac{-\lambda_{s}}{\mu_S}
\left(  V_1^a V_2^b + V_2^a V_1^b  \right),
~~~
\label{solS} \\
\left< {\cal A}^{ab} \right> &=& \frac{\lambda_a}{\mu_A}
\left(  V_1^a V_2^b -  V_2^a V_1^b  \right),
\label{solA}
\end{eqnarray}
and analogously for $\left< {\bar{\cal S}}_{ab} \right>$ and $\left< \bar{{\cal A}}_{ab} \right>$.
Here $V_1^a = \left< {\cal F}_1^a \right>$ and  $V_2^b = \left< {\cal F}_2^b \right>$.
We can always give an explicit expression for the minimum in a particular U(2) basis where
the vevs $V_1$ and $V_2$ adopt the form,
\begin{equation}
V_1 = 
\left[
\begin{array}{c}
v e^{i\psi} \\
- v_1 e^{i\phi} 
\end{array}
\right], \quad
V_2 = 
\left[
\begin{array}{c}
0 \\
v_2 e^{- i\phi^{\prime}} 
\end{array}
\right].
\label{U2basis} 
\end{equation}
Here the vevs $v,v_{1}$ and $v_{2}$ are defined positive and 
the sign in the entry $V_{1}^{2}$ has been introduced for convenience
in the rest of the analysis.
Since the solutions given by Eqs.~\ref{solS} and \ref{solA} cancel the 
${\cal V}_S$ and ${\cal V}_A$ components of the scalar potential, we must require
the remaining component of the scalar potential, ${\cal V}_F$, to cancel
if the solution is to be a minimum. For instance, using Eqs.~\ref{solS} and \ref{solA},
the component $\left| \partial {\cal W} / \partial {\cal F}_{1}\right|^{2}$ of ${\cal V}_{F}$ 
can be written as, 
\begin{equation}
\left| \frac{ \partial {\cal W} }{ \partial {\cal F}_{1} }\right|^{2}
= \rho_1^{2} {\cal F}_1 \bar{\cal F}_1 +
\rho_3^{2} {\cal F}_2 \bar{\cal F}_2 +
\rho_{1} \rho_{3} (  {\cal F}_1 \bar{\cal F}_2 +   {\cal F}_2 \bar{\cal F}_1 ) -
\rho_{1} \frac{\lambda_{s}^{2}}{\mu_{s}} {\cal F}_1 \bar{\cal F}_2   {\cal F}_2 \bar{\cal F}_1 
+ \cdot \cdot \cdot
\label{vfeq}
\end{equation}
We are interested in finding at least one solution consistent with the data. 
Let us assume that the bilinear terms in ${\cal V}_{F}$ dominate. Later we will see
that this assumption is consistent with $\mu_{S} \ll \rho_{i}$.
In the basis given by Eqs.~\ref{U2basis} this condition translates into
the equation,
\begin{equation}
\mu_1 v^2 + ( \sqrt{\mu_1} v_1 - \sqrt{\mu_2} v_2)^2 +
2 ( \sqrt{\mu_1 \mu_2} - \mu \cos (\phi+\phi^{\prime}) ) v_1 v_2 =0.
\end{equation}
where $\mu_{1} =( \rho_{1}^{2} + \rho_{4}^{2})$, 
$\mu_{2} =( \rho_{2}^{2} + \rho_{3}^{2})$ and 
$\mu =( \rho_{1}\rho_{3} + \rho_{2} \rho_{4})$.  
This equation admits a non-trivial vacuum given by
\begin{eqnarray}
v_2  &=& \sqrt{\frac{\mu_1}{\mu_2}} v_1, \\
\cos (\phi + \phi^{\prime} ) &=& 
\frac{ \sqrt{\mu_1 \mu_2}}{\mu}\left( 1  - \frac{ v^2}{2v_1^{2}} \right).
\label{alphaEq1} 
\end{eqnarray}
We note that the existence of a non-trivial vacuum requires the presence
of both vectorial Higgs fields, ${\cal F}_{1}$ and ${\cal F}_{2}$. 
If one of these fields were not present, it would not be possible to break
the flavor symmetry at the supersymmetric level, which would force us 
to resort to supersymmetry breaking terms \cite{Barbieri:1998em}.
\section{SU(5) representations of the Higgs fields}
Yukawa couplings for the first and second generations are generated at higher order
by non-renormalizable interactions that are generically of the form
\begin{eqnarray}
\frac{1}{M} \left(  \Psi_{10} {\cal S} \Psi_{10} {\cal H}_5 +
 \Psi_{10} {\cal A} \Psi_{10} {\cal H}_5 + \right.
\nonumber  \\
\left. +  \Psi_{\bar{5}} {\cal S} \Psi_{10} {\cal H}_{\bar{5}} +
 \Psi_{\bar{5}} {\cal A} \Psi_{10} {\cal H}_{\bar{5}} \right),
\end{eqnarray}
where flavor indices have been omitted.
Yukawa couplings mixing the first and second generations 
with the third generation are also generated at higher order
by non-renormalizable interactions,  
\begin{equation}
\frac{1}{M}\left(  \sigma_u \psi_{10} {\cal F}_j \Psi_{10} {\cal H}_5 +
 \psi_{10} {\cal F}_j \Psi_{\bar{5}} {\cal H}_{\bar 5} + 
 \psi_{\bar{5}} {\cal F}_j \Psi_{10} {\cal H}_{\bar{5}} \right)
\label{operi3}
\end{equation}
where $j=1,2$ and
again flavor indices have been omitted.
Here we find it convenient to introduce an 
additional real coupling, $\sigma_u$,
to differentiate the relative size and sign of the contributions
to the (13) and (23) mixing in the down and up-type quark sectors.
If the flavor breaking Higgs fields were SU(5) singlets, then,
after flavor symmetry breaking, $3\times 3$ symmetric  
Yukawa matrices would be generated for the charged leptons 
and down-type quark fields, of the form,
\begin{equation}
{\cal Y} = 
\left[
\begin{array}{cc}
\frac{1}{M} ( \left< {\cal S} \right> + \left< {\cal A} \right> ) &
\frac{1}{M} ( \left< {\cal F}_1 \right> + \left< {\cal F}_2 \right> )  \\
\frac{1}{M} ( \left< {\cal F}_1 \right> + \left< {\cal F}_2 \right> )  &
h
\end{array}
\right],
\label{genYuk}
\end{equation}
where $h$ represents generically a third generation Yukawa coupling.
This Yukawa matrix, nevertheless, would predict
wrong mass relations between the charged lepton and the down-type quark sector,
$m_{e}/m_{\mu} = m_{d}/m_{s}$ and $m_{\mu}/m_{\tau} = m_{s}/m_{b}$ .

It was pointed out some time ago that it may be possible
to explain some of the observed differences between the charged lepton and
down-type quark sectors by
promoting the SM vertical symmetry to a GUT symmetry as
the SU(5) symmetry of Georgi and Glashow \cite{Georgi:1974sy}. 
In the context of SU(5), the observed 
empirical relations find one on their simplest explanations,
the so-called Georgi-Jarlskog mechanism \cite{Georgi:1979df}, which 
in the case of a U(2) flavor symmetry 
adopts a particular implementation proposed in Ref.~\cite{Barbieri:1996ww}.
If the flavor-symmetric field ${\cal S}$ transforms as a ${\bf 75}$ representation
of SU(5) the tensor product, ${\cal S}{\cal H}_{\bar 5}$ transforms effectively as a 
$\bar{ \bf 45}$ representation of SU(5). 
It is known that this would account perfectly for the approximate
empirical factor 3 that contects the muon-tau and 
strange-bottom mass ratios (at the GUT scale), $m_{\mu}/m_{\tau} \approx 3 m_{s}/m_{b}$.  
Furthermore if the flavor-antisymmetric tensor ${\cal A}$
transforms as a SU(5) singlet, the operator 
$ \Psi_{\bar{5}} {\cal A} \Psi_{10} {\cal H}_{\bar{5}}$
will generate the same contributions to the (12) and (21) entries
in both the charged lepton and down-type quark Yukawa matrices.
This, together with the representation properties of ${\cal S}$ postulated above,
could help, as we will see later, to explain a second well known empirical relation connecting
the down-strange and the electron-muon mass ratios, 
$(m_{d}/m_{s})^{1/2} \approx 3 (m_{e}/m_{\mu})^{1/2}$.
Finally the vector fields ${\cal F}_1$ and ${\cal F}_2$ could transform
as a singlet or alternatively under the representation {\bf 24} of SU(5). In both cases
the operators in Eq.~\ref{operi3} would generate the entries (i3) 
and (3i), (i=1,2) of the Yukawa matrices. 
\section{The down-type quark Yukawa matrix}
We note that 
the Yukawa matrices generated by the model
could in principle contain complex phases in all the entries.
Some of these phases do not  generate observable effects 
in the fermion sector. In this model one can always 
redefine the phases of the matter fields
to make the upper left submatrix real,
which is just a convenient basis to
study CP-violating effects.
This phase redefinition, which does not
change the predictions of the model, can be simply enforced assuming that the entries
(12),(21) and (22) of the
vevs $\left<{\cal S}\right>$ and $\left<{\cal A}\right>$ are real. 
In doing so we obtain relations between the phases of Higgs fields
that allow us to see more clearly the predicitions that the model makes
for the CKM phase. 
In the U(2) basis given by Eq.~\ref{U2basis}
the vev $\left<{\cal A}\right>$ takes the form,
\begin{equation}
\left< {\cal A} \right> =
\frac{\lambda_a e^{i (\psi -\phi^{\prime})}}{\mu_A}  
\left[
\begin{array}{cc}
0 & 1 \\
-1 & 0 
\end{array}
\right].
\end{equation}
The condition ${\rm Im} \left< {\cal A} \right> =0$ implies the relation
$\psi= \phi^{\prime}$. In the same basis, and assuming $\psi= \phi^{\prime}$,
${\rm Im} \left< {\cal S} \right>$ is given by,
\begin{equation}
\frac{-2}{\mu_S}  
\left[
\begin{array}{cc}
0  & 0 \\
0 &  \lambda_s v_1 v_2 \sin (\phi - \phi^{\prime})
\end{array}
\right] . 
\end{equation}
The condition ${\rm Im} \left< {\cal S} \right>=0$ implies that the 
phases $\phi$ and $\phi^{\prime}$ are equal, 
$\phi^{\prime}=\phi$, with $\phi$ undetermined.
Taking this into account we can adopt, for convenience, 
an equivalent alternative notation
for the vevs $V_1$ and $V_2$,
\begin{equation}
V_1 = 
\left[
\begin{array}{c}
v e^{i\frac{\alpha^{\prime}}{2}} \\
- v_1 e^{i\frac{\alpha^{\prime}}{2}} 
\end{array}
\right], \quad
V_2 = 
\left[
\begin{array}{c}
0 \\
v_2 e^{- i \frac{\alpha^{\prime}}{2}} 
\end{array}
\right].
\label{U2basisalpha} 
\end{equation}
The phase $\alpha^{\prime}$ will be the only phase that manifests in the Yukawa matrices.
A non-zero value would signal the appeareance of CP-violation. 
Using this notation Eq.~\ref{alphaEq1} can be rewritten as,
\begin{equation}
\cos (\alpha^{\prime} ) = 
 \frac{\sqrt{ \mu_1 \mu_2}}{\mu}\left( 1 - \frac{ v^2}{2v_1^{2} }\right). 
\label{alphaEq2}
\end{equation}
The down-type quark Yukawa matrix 
can be conveniently rewritten in the form,
\begin{equation}
{\cal Y}_D = h_b
\left[
\begin{array}{ccc}
0 &  a v \lambda^+ & 2  a \zeta \eta e^{i \frac{\alpha^{\prime}}{2}} \\
 a v \lambda^- & 
2 a v_1 \widehat{\lambda}_{s} 
&   a \lambda_{cb} \\
2  a \zeta \eta  e^{i \frac{\alpha^{\prime}}{2}} &  a \lambda_{cb} & 1
\end{array}
\right].
\label{YD}
\end{equation}
where,
\begin{eqnarray}
\lambda^{\pm} &=& 
 - \left( \widehat{\lambda}_s \mp \widehat{\lambda}_a \right) , \\
\widehat{\lambda}_{s} &=&  \frac{\lambda_s}{\mu_S},
\quad
\widehat{\lambda}_{a} = \frac{\lambda_a}{\mu_A},
 \\
\lambda_{cb} &=&  ( e^{-i \frac{\alpha^{\prime}}{2}} -
\zeta e^{i \frac{\alpha^{\prime}}{2}} ) 
\end{eqnarray}
and 
\begin{equation}
\zeta = \frac{v_1}{v_2}, \quad
a = \frac{v_2}{M}, \quad \eta = \frac{v}{2 v_1}.
\label{alz}
\end{equation}
The parameters $a$, $\eta$ and $\zeta$, as we will analyze later, are 
directly correlated with the absolute values of the CKM elements. 
Therefore we naively expect them to be smaller than 1. 
In Secs.~\ref{fit} and \ref{preds} we will study the precision predictions
of the model. For the moment, let us assume that $ 2\eta \zeta \lesssim v \lambda^{\pm} $.
In that case the down-strange and electron-muon mass ratios 
predicted by the model are related. These are given by,
\begin{equation}
\frac{m_d}{m_s} \approx \eta^2 
\frac{ ( \widehat{\lambda}_s^2 -\widehat{\lambda}_a^2 )}{\widehat{\lambda}_s^2 },
\quad 
\frac{m_e}{m_{\mu}} \approx \eta^2 
\frac{ ( 9 \widehat{\lambda}_s^2 -\widehat{\lambda}_a^2 )}{9\widehat{\lambda}_s^2 }.
\label{mdms}
\end{equation}
\section{The up-type quark Yukawa matrix}
Assuming that the U(2) flavor breaking fields 
${\cal S}$, ${\cal A}$ and ${\cal F}_i$ transform under the representations
{\bf 75}, {\bf 1} and {\bf 1}, respectively, implies that two of the associated
higher order operators in the up-type quark sector are exactly zero,
\begin{equation}
 \Psi_{10} {\cal S} \Psi_{10} {\cal H}_5 =0, \quad
 \Psi_{10} {\cal A} \Psi_{10} {\cal H}_5 =0.
\end{equation}
If this were the case the up-type quark Yukawa matrix would have the form,
\begin{equation}
{\cal Y}_U = h_t
\left[
\begin{array}{ccc}
0 & 0 & 2 \theta_u a \zeta \eta e^{i \frac{\alpha^{\prime}}{2}} \\
0 & 0 &  \theta_u a  \lambda_{cb} \\
2 \theta_u a \zeta \eta  e^{i \frac{\alpha^{\prime}}{2}} 
& \theta_u a \lambda_{cb} & 1
\end{array}
\right].
\end{equation}
where $\theta_u = \sigma_u /h_t$.
This matrix is afflicted by a serious phenomenological problem:
it predicts that the up-quark is massless.  
Although the possibility of a massless up-quark has been considered 
in the past as a solution to the strong CP-problem, more recent 
studies of pseudoscalar masses and decay constants, along with
other arguments, suggest that the up-quark mass is nonzero.
Other flavor models proposed in the literature, when faced with the problem
of the smallness of the up-quark mass, have resorted to introducing
additional Higgs fields. Interestingly in the current model one can generate
a mass for the up quark at second order in the effective operator expansion
without resorting to additional Higgses.
At second order in powers of $1/M$ there are only three terms
that can contribute to the up-type quark Yukawa matrix.  
These terms are generically of the form,
\begin{equation}
\frac{1}{M^2}  \Psi_{10} \Psi_{10} {\cal H}_5 
\left( \lambda^{\prime}  {\cal F}_i  {\cal F}_i  +
\lambda^{\prime \prime}  {\cal F}_1  {\cal F}_2 
\right),
\label{opersec}
\end{equation}
where $i,j=1,2$ and flavor indices have been omitted. 
The two terms proportional to $\lambda^{\prime}$ will generate an
additional contribution to the charm quark mass but they will not contribute to
the up-quark mass. The up-quark mass will be generated by 
the operator proportional to $\lambda^{\prime \prime}$.
Taking all three contributions into account
the up-type quark Yukawa matrix takes the form,
\begin{equation}
{\cal Y}_U = h_t
\left[
\begin{array}{ccc}
 a^2 \frac{\lambda^{\prime} v^2 }{v^2_2} & 
 a^2 \lambda_{uc}  \frac{v}{v_2}  
 & 2 \theta_u a \zeta \eta e^{i \frac{\alpha^{\prime}}{2}} \\ 
 a^2 \lambda_{uc} \frac{v}{v_2}  
& a^2 \lambda_{ct} 
& \theta_u a \lambda_{cb} \\
2 \theta_u a \zeta \eta  e^{i \frac{\alpha^{\prime}}{2}} 
& \theta_u a \lambda_{cb} & 1
\end{array}
\right],
\end{equation}
where,
\begin{eqnarray}
\lambda_{ct} &=& ( \lambda^{\prime \prime} \zeta +\lambda^{\prime} \zeta^2 
+ \lambda^{\prime}), \\
\lambda_{uc} &=&  (\lambda^{\prime \prime} +\lambda^{\prime} \zeta ).
\end{eqnarray}
The parameters $a$, $\eta$ and $\zeta$ were defined in Eq.~\ref{alz}.
\section{Determination of the parameters of the model \label{fit}}
In this section we will analyze 
how to determine the parameters of the model using 
some of the quark mass ratios and mixing angles.
In the next section we will study the predictions
the model makes once its parameters have been determined. 
We will make our point by proving that there is at least 
one particular limit of the parameter space
that can reproduce observations.
We will see that at this point of the parameter space 
the fundamental parameters are easily calculable. 
Let us assume that the following hierarchies
in the lagrangian parameters hold, 
\begin{eqnarray}
\theta_u &\ll& 1, \\
\frac{\lambda_a}{\mu_A} &\approx &  - 2\frac{\lambda_s}{\mu_S}, \\
\sqrt{\mu_1 \mu_2} &\ll& \mu,  \\ 
 \lambda^{\prime\prime} &\ll& \lambda^{\prime}.
\end{eqnarray}
The down and up-type quark Yukawa matrices can be brought to
diagonal form by a biunitary diagonalization, 
$({\cal V}_L^d)^{\dagger} {\cal Y}_D {\cal V}_R^d=
(h_d,h_s,h_b)$ and $({\cal V}_L^u)^{\dagger} {\cal Y}_U {\cal V}_R^u=(h_u,h_c,h_t)$. 
The CKM matrix is defined by ${\cal V}_{\rm CKM} = ({\cal V}_L^u)^{\dagger} {\cal V}_L^d$.
If $\theta_u \ll 1$ the CKM elements are to leading order determined
from the mixing in the down-type quark Yukawa matrix, {\it i.e.}  ${\cal V}_{\rm CKM} \approx {\cal V}_L^d$,
while the mixing arising from the up-type quark sector can be neglected.
Therefore the unitary matrix
${\cal V}_{\rm CKM}$ can be calculated from Eq.~\ref{YD}. 
To leading order it is a function of the four real parameters $\zeta$,$a$,$\lambda$, 
$\eta$ and the complex phase $\alpha^{\prime}$ 
\begin{equation}
{\cal V}_{\rm CKM} \approx  
\left[
\begin{array}{ccc}
1 -  \frac{\lambda^{2}}{2}  &  \lambda &  2 \eta \zeta  a  e^{ i \alpha^{\prime}/2} \\
\lambda  &  \frac{\lambda^{2}}{2} +\frac{a^{2}r^{2}}{2} -1  &   (e^{ -i \frac{\alpha^{\prime}}{2}} - \zeta e^{ i \frac{\alpha^{\prime}}{2}})  a \\  - a \lambda ( e^{ i \frac{\alpha^{\prime}}{2}} - \zeta ( 1 - 2 \eta /\lambda ) e^{-i \frac{\alpha^{\prime}}{2}})     
&    (  e^{ i \frac{\alpha^{\prime}}{2}} - \zeta e^{ -i \frac{\alpha^{\prime}}{2}}) a  & 1- \frac{a^{2 }r^{2}}{2}
\end{array}
\right].
\label{CKMCP}
\end{equation}
where $r^{2} =  ( 1 - 4 \zeta \cos\alpha^{\prime} +4 \zeta^{2} )$. Using the second assumption in our
parameter space,  $\widehat{\lambda}_a \approx -2  \widehat{\lambda}_s$, we obtain from Eq.~\ref{YD},
\begin{equation}
\lambda = - \eta \frac{ (\widehat{\lambda}_s + \widehat{\lambda}_a)} { \widehat{\lambda}_s} \approx \eta.
\end{equation}
The parameters $\lambda$, $a$ and $\zeta$
can be determined using experimental data.
The parameter $\lambda$ is to leading order given by $\lambda = \left| V_{us} \right|$.
The parameters $a$ and $\zeta$ can be determined using 
the absolute values of $\left| V_{us} \right|$, $ \left| V_{ub} \right|$
and $ \left| V_{cb} \right|$ by,
\begin{equation}
a  = \frac{ \left| V_{cb} \right|}{r}, \quad 
\frac{\zeta}{r}=
 \frac{  \left|  V_{ub} \right| }{  2 \left| V_{cb}  \right| \left|V_{us} \right|  } .
\label{zeta}
\end{equation}
It is a trivial check to prove that
the angle $\alpha^{\prime}$ introduced in the parametrization of
the CKM matrix given in Eq. \ref{CKMCP} coincides to leading order
in powers of $\zeta$ with the standard definition of $\alpha$, 
$$
\alpha = {\rm Arg} \left[ -\frac{ V_{td}V^{*}_{tb}}{  V_{ud}V^{*}_{ub}} \right]
= \alpha^{\prime} - \zeta.
$$
Furthermore, with our third assumption on the parameter space of the model,
$\sqrt{\mu_1 \mu_2}\ll \mu$, Eq.~\ref{alphaEq2} predicts 
the angle $\alpha^{\prime}$ to be $\pi/2$ since
\begin{equation}
\cos \alpha^{\prime} \propto \frac{\sqrt{\mu_{1} \mu_{2}}}{\mu} \ll 1 .
\end{equation}
Therefore the phase $\alpha$ in that limit of the parameter space is given by
\begin{equation}
\alpha \approx \frac{\pi}{2} -\zeta . 
\end{equation}
Assuming that $\alpha^{\prime}\approx \pi/2$ and using Eqs.~\ref{zeta}
the parameters $a$ and $\zeta$, as a good approximation, are given by,
\begin{equation}
a \approx \left| V_{cb} \right| , \quad 
 \zeta \approx 
\frac{  \left| V_{ub} \right| }{ 2 \left| V_{cb} \right| \left| V_{us} \right|}
\end{equation}
Using the values of the CKM elements given by the PDG collaboration \cite{Eidelman:2004wy},
$\left| V_{us} \right|= 0.220 \pm 0.0026$, 
 $\left| V_{ub} \right|= 0.00367 \pm 0.00047$
and $\left| V_{cb} \right|= 0.0413 \pm 0.0015$
and the ratio $\left| V_{ub}\right|/\left| V_{cb}\right| = 0.086\pm 0.008$ \cite{Battaglia:2003in}, 
we obtain approximately $\zeta \approx 0.19 \approx \lambda$ and $a\approx \lambda^2$.  
The couplings $\lambda^{\prime}$ and $\lambda^{\prime \prime}$ can be 
determined from the up-charm and charm-top quark mass ratios.
Taking into account that $a,\lambda, \zeta \ll 1$
and making use of our fourth assumption on the parameter
space of the model, $\lambda^{\prime \prime} < \lambda^{\prime}$, 
these are given to leading order by,
\begin{eqnarray}
\frac{m_c}{m_t} &=&  a^2 \lambda^{\prime} ( 1 + \zeta \rho + \zeta^2 ), \\
\frac{m_u}{m_c} &=&  \frac{ 8 \lambda^2 \zeta^3 \rho }{(\zeta + \rho)}. 
\end{eqnarray}
Here $\rho = \lambda^{\prime \prime} /\lambda^{\prime}$.
Numerically, as a good approximation, 
$m_c/m_t \approx \lambda^3 /2$ and $m_u/m_c \approx \lambda^4$
(at low energy) \cite{Ferrandis:2004ti}. 
Therefore, taking into account that $a\approx \lambda^{2}$,
a good estimation of the couplings $\lambda^{\prime}$ and $\lambda^{\prime \prime}$ is
$\lambda^{\prime} \approx 3$ and $\lambda^{\prime\prime} \approx \lambda \lambda^{\prime}$.
Finally there is one more dimensionless model parameter, $ v_{1} \widehat{\lambda}_{s}$, 
that appears in the down-quark type Yukawa matrix, ${\cal Y}_D$,
and that has to be determined from the strange-bottom mass ratio,
\begin{equation}
\frac{m_s}{m_b} =  2 a v_{1} \widehat{\lambda}_s. 
\end{equation}
Numerically the ratio $m_s/m_b$ is know to be approximately $\lambda^2 /2$ (at low energy)  
\cite{Ferrandis:2004ti}. 
Therefore 
$$
v_1 \widehat{\lambda}_s  \approx \lambda.
$$
Taking this into account we are now ready to understand that neglecting
quartic terms in Eq.~\ref{vfeq} was perfectly consistent with the
assumption $\mu_{s} \ll \rho_{i}$.
We note that the numerical estimations of the parameters of the model 
have been implemented at the scale of flavor breaking, 
which we assumed to be close to the GUT scale. 
The parameters $\lambda$, $\zeta$ and $\rho$ are very 
approximately renormalization scale independent.
On the other hand the parameters $a$, $\lambda^{\prime}$ and $v_1 \widehat{\lambda}_s$,
which are determined from $\left| V_{cb}\right|$, $m_c/m_t$ and $m_s/m_b$ respectively,
depend on the location of the flavor breaking scale. 
As a consequence, renormalization corrections to these observables, whose expressions
are available in the literature
\cite{Ma:1979cw,Babu:1992qn}, must be implemented if a precision calculation of
the fundamental parameters of the model is required.
\section{Predictions \label{preds}}
Once the fundamental parameters of the model have been determined 
as in the previous section, we are able to make three real
predictions for fermion mass ratios. First we note that
the model predicts $({\cal Y}_D)_{11}=0$, while $({\cal Y}_D)_{13}$,
which is determined from $\left| V_{ub} \right|$, plays a minor role in the determination
of the down quark mass. We thus obtain the prediction,
\begin{equation}
\frac{m_d}{m_s} \approx  \left| V_{us} \right|^2.
\end{equation}
This is a prediction that is well confirmed by the data.
It is also an empirical relation that has been known for 37 years \cite{Gatto:1968ss}. 
Using the sum rules to extract the masses of the lighter quarks we obtain,
$(m_d /m_s)^{1/2} = 0.209\pm 0.019$ \cite{Ferrandis:2004ti}. 

In the charged-lepton sector, given the proposed SU(5) representation assignments of
the Higgs fields, our model predicts that the charged lepton Yukawa matrix,
${\cal Y}_L$, is the same as the down-type quark Yukawa matrix, except that the contributions
from the symmetric tensor ${\cal S}$ contain an additional factor 3.
We thus obtain two predictions for the charged lepton mass ratios (see Eqs.~\ref{mdms}), 
\begin{eqnarray}
\left( \frac{m_e}{m_{\mu}}\right)^{1/2} &\approx & \frac{1}{3} \left( \frac{5}{3} \right)^{1/2}
\left( \frac{m_d}{m_s}\right)^{1/2} \approx  \frac{1}{2.5} \left( \frac{m_d}{m_s}\right)^{1/2}  , \\
\left( \frac{m_{\mu}}{m_{\tau}}\right) & \approx & 3
\left( \frac{m_s}{m_b}\right) , 
\end{eqnarray}
The first of these predictions, which is approximately renormalization scale independent,
is fully consistent at $1\sigma$ with the experimental data \cite{Ferrandis:2004ti}, 
\begin{equation}
\left. \frac{ (m_d / m_s)^{1/2}}{ (m_e /m_{\mu})^{1/2}} \right. _{\rm exp}  =  3.06 \pm 0.48.
\end{equation}
The second prediction is renormalization scale dependent
and holds at the flavor breaking scale.
Using experimental values we obtain at the electroweak scale  \cite{Ferrandis:2004ti},
\begin{equation}
\left. \frac{ (m_{\mu} / m_{\tau}) }{ (m_s /m_{b})}  \right. _{\rm exp}= 
 2.55 \pm 0.4 . 
\end{equation}
Extrapolating to the GUT scale,
$10^{16}$~GeV, in a supersymmetric scenario (with low $\tan\beta$) we obtain,
$(m_{\mu}/m_{\tau})_{\rm GUT} = ( 3.14 \pm 0.4 )\times (m_{s}/m_{b})_{\rm GUT} $.

Finally we will show that the prediction that the model makes in the limit 
$\sqrt{\mu_{1} \mu_{2}} \ll \mu$ for the phase $\alpha^{\prime}$ 
successfully reproduces the measured value of CP violation in the quark sector.
We have shown in the previous section that the model contains a particular
limit, $\sqrt{\mu_1 \mu_2} \ll \mu$, where the CP phase $\alpha$ is predicted
to be close to $\pi/2$. It is convenient to define the variables,
\begin{equation}
\delta^{\prime} = \frac{\pi}{2} - \alpha^{\prime}, \quad
\delta = \frac{\pi}{2} - \alpha,  
\label{delta}
\end{equation}
and solve the Eq.~\ref{zeta} for $\zeta$ expanding 
around $\delta^{\prime}=0$. We obtain,
\begin{equation}
\zeta = \frac{1}{\sqrt{3}}  ( 1 - \frac{4}{3} z) + \frac{\delta^{\prime}}{3} ( 1 - \frac{8}{3} z) 
\label{azetasol},
\end{equation}
where $z$ is a perturbative parameter defined as,
\begin{equation}
z = 1 -  \frac{ 2 \left|  V_{ub} \right| }{   \left| V_{cb}  \right| \left|V_{us} \right|  }
\label{zeq}.
\end{equation}
$z$ can be determined from the measured absolute values of the
CKM elements. Using the value of $\left| V_{us} \right|$ given by the PDG collaboration
and the ratio $\left| V_{ub}\right|/\left| V_{cb}\right| = 0.086\pm 0.008$ \cite{Battaglia:2003in},
we obtain $z = 0.22 \pm 0.08$. 
Finally using our parametrization of the CKM matrix
we obtain a simple expression for the leading-order relation between the angles $\beta$ and 
$\alpha$, 
\begin{equation}
\beta
= {\rm Arg} \left[ - \frac{V_{cd}V^{*}_{cb}} {V_{td}V^{*}_{tb}} \right]
=  {\rm Arg} \left[ 1 -  \zeta e^{-i \alpha}\right].
\label{betaalpha}
\end{equation}
Using the formula for $\zeta$ given in 
Eq.~\ref{azetasol} and expanding around $z,\zeta,\delta=0$  we find that 
in the limit $\sqrt{\mu_1 \mu_2}  \ll \mu$, the model predicts 
the phase $\beta$ to be, 
\begin{equation}
\beta
= \frac{\pi}{6} + \frac{\delta}{2} - \frac{ z}{\sqrt{3}} + {\cal O}(\delta^{2}).
\label{betaz}
\end{equation}
For instance, 
using the numerical values of $z$ and taking into account that 
$\delta \approx \zeta$, 
as calculated above from the absolute values of the CKM elements,
we find that,
$\beta = 28.5^{\circ} \pm 5 ^{\circ}$,
while $\gamma \approx \pi/2 - \beta + \delta = 76.5^{\circ} \mp 5^{\circ}$.
These values are within the 1 sigma windows for the 
measured value for $\beta$, $\beta_{\rm exp} = 23.3^{\circ} \pm 1.6^{\circ}$
\cite{betaexp,Ferrandis:2005jb} and the indirect determination through CKM global fits for $\gamma$,
\cite{fits}, $\gamma_{\rm fit} \approx 61^{\circ} \pm 11^{\circ}$
at $95\%$ C.L..

Some comments must be added regarding the size of the indirect effects of new physics in flavor changing and CP processes.  All these processes appear usually in many susy models as a consequence of the presence of flavor and CP violating terms in the soft supersymmetry breaking sector. In many susy flavor models the breaking of flavor symmetries directly generates those terms in the soft susy breaking sector. In this model this is not so because the flavor symmetry is broken "before" supersymmetry is broken. Although the mechanism of susy breaking has not been specified this is not so relevant since, for instance, two different alternative mechanisms could explain the suppression of those terms:
A)  flavor is broken at the supersymmetric level of the theory at the GUT scale. The supersymmetry is broken at a much lower scale that could go from around $10^{11}$~GeV to $10^{4}$~ GeV. The radiative transmission of flavor violation to the susy breaking sector would be suppressed in this case because it would require the presence in the loops of particles of very disparate scales. If for instance the flavor symmetry were a local gauge symmetry it would require the presence in the loops of flavored Higgses or flavored gauge bosons, whose masses are of the order of the GUT scale, together with susy particles whose masses are much lower, B)  alternatively supersymmetry could be broken through a flavor blind mechanism as in for instance gauge mediated models and this problem would not be an issue. 
\section{Conclusions}
We have proposed a supersymmetric grand unified model 
for the breaking of flavor and CP symmetries. The model fits
all the data in the charged lepton sector with precision and 
makes three successful predictions for fermion mass ratios.
We believe that this model has the following features 
worth emphasizing,
\begin{itemize}
\item
the model allows us to solve the flavor scalar potential analytically,
\item
supersymmetry is the key ingredient that makes the model predictive
and solvable,
\item
the presence of two vector flavored Higgses is necessary for the 
existence of a non-trivial supersymmetric vacuum,
\item
all the Higgses introduced play a role in the breaking of the flavor symmetry,
\item
no additional Higgses have to be postulated to generate a mass for the up-quark,
\item
and, most importantly, in this model CP symmetry is broken in 
a predictive way.
\end{itemize}
We should point out that many models for the spontaneous breaking
of CP are more descriptive than predictive. Certain parameters must be tuned
so as to be able to reproduce the data. 
On the other hand, the model here proposed contains a simple limit of its 
parameter space where the mesured $\beta$ phase is successfully reproduced.
In this regard, a precise measurement of $\gamma$ and $\alpha$,
which is the next challenge for B factories, is of paramount importance.
It is expected \cite{futuregamma,superB} that if the luminosity in the upgraded B 
factories is high enough, $\gamma$ could be determined 
with approximately $5\%$ percent uncertainty; a measurement
with an uncertainty at the level of $1\%$
will require a superB factory.
A determination of $\alpha$ at less than the $5\%$ level
may also have  to wait for the superB factories \cite{superB}.   
\acknowledgements
I thank H.~Guler for many suggestions.
This work is supported by:
the Ministry of Science of Spain under grant EX2004-0238,
the D.O.E. under Contracts: DE-AC03-76SF00098 and 
DE-FG03-91ER-40676 and
by the N.S.F. under grant PHY-0098840.

\end{document}